\documentstyle[12pt]{article}
 
\begin{document}

\baselineskip=14pt plus 0.2pt minus 0.2pt
\lineskip=14pt plus 0.2pt minus 0.2pt

\begin{flushright}
quant-ph/9612050 \\
LA-UR-96-4789 \\
\end{flushright}

\begin{center}
\large{\bf Displaced and Squeezed Number States} 

\vspace{0.25in}

\bigskip

 Michael Martin Nieto\footnote
{Email: mmn@pion.lanl.gov }

\vspace{0.3in}

{\it
Theoretical Division\\
Los Alamos National Laboratory\\
University of California\\
Los Alamos, New Mexico 87545, U.S.A.\\
\vspace{.25in}
and\\
\vspace{.25in}
Abteilung f\"ur Quantenphysik\\
Universit\"at Ulm\\
D-89069 Ulm, GERMANY\\}

\vspace{0.3in}

{ABSTRACT}

\end{center}

\vspace{0.25in}

\baselineskip=0.333in

\begin{quotation}
After beginning with a short historical review of the concept of displaced 
(coherent) and squeezed states, we discuss previous (often forgotten) work on
displaced and squeezed number states.  Next, we obtain the most general
displaced and squeezed number states.  We do this in both the functional and 
operator (Fock) formalisms, thereby demonstrating the necessary equivalence. 
We then obtain the time-dependent expectation values,  uncertainties, 
wave-functions, and probability densities.  In conclusion, there is a 
discussion on the possibility of experimentally observing these states.

\end{quotation}

\vspace{0.3in}

\newpage

\baselineskip=.33in

\section{Background}

The coherent states were discovered in 1926 by Schr\"odinger
\cite{sch}, as an 
example of how his wave functions could mimic classical particles.
They consisted of Gaussians in an harmonic oscillator 
potential.  The widths of these Gaussians were those of the 
ground-state Gaussian.  In modern notation, they were 
\begin{equation}
\psi = \pi^{-1/4}\exp\left[-\frac{(x-x_0)^2}{2}+ip_0x\right]~.\label{sch}
\end{equation}
It is to be noted that, this being before Born's
probability interpretation of $\psi^*\psi$, Schr\"odinger was 
concerned by the complex nature of the wave function, wondering if it was 
the real part that was significant. 

Lorentz was the person who had been 
especially bothered by the lack of classical 
properties of eigenstate wave functions.    
Schr\"odinger and Lorentz exchanged many letters on this subject, and their 
correspondence is published \cite{lor}.  

Soon after this, Kennard \cite{kennard} published a paper on quantum 
motion.   There he described what are,  
in modern parlance, squeezed states.  They follow
the classical motion, they are Gaussians whose widths are not that 
of the ground state, and the widths and uncertainty products 
oscillate with time.  However, relatively little notice was given to this 
paper in later times \cite{saxon}.

In the 1960's the emergence of quantum optics was a fertile 
background for the 
modern development of coherent states by Glauber,
Klauder, and Sudarshan \cite{glauber}-\cite{ks}.  Using boson operator 
calculus, these states could be defined as either the eigenstates 
of the destruction operator, 
\begin{equation}
a|\alpha\rangle = \alpha|\alpha\rangle ,
\end{equation}
or equivalently as the state obtained by operating on the ground 
state by the displacement operator:
\begin{equation}
D(\alpha)|0\rangle = |\alpha\rangle~,
\end{equation}
\begin{equation}
D(\alpha)= \exp[\alpha a^{\dagger} - \alpha^*a]
     = \exp[-|\alpha|^2/2]\exp[\alpha a^{\dagger}] \exp[-\alpha^* a]~,
\end{equation} 
the last equality coming from a Baker-Campbell-Hausdorff 
(BCH) calculation.  
These states are 
\begin{equation}
|\alpha\rangle = e^{-|\alpha|^2/2}\sum_{n=0}^{\infty}
           \frac{\alpha^n}{\sqrt{n!}} |n\rangle ~.
\end{equation}
Use of a generating formula for Hermite polynomials shows that this 
yields the same wave function as Eq. (\ref{sch}), with the identifications
\begin{equation}
\alpha=\alpha_1+i\alpha_2= \frac{x_0+ip_0}{\sqrt{2}}~.
\end{equation}

Squeezed states were rediscovered and elucidated by 
a number of people \cite{many}-\cite{ss3}.  (Ref. \cite{ss2} was where 
the term ``squeeze" was invented.  See Ref. \cite{nato} for a
discussion of this history.)  In operator form, these states are 
created by 
\begin{equation}
D(\alpha)S(z)|0\rangle = |\alpha,z\rangle~, \label{xx}
\end{equation}
where the squeeze operator is 
\begin{equation}
S(z) = \exp\left[\frac{1}{2}za^{\dagger}a^{\dagger}-\frac{1}{2}z^*aa\right]~,
 ~~~~~~  z = re^{i\phi} = z_1 + i z_2~. 
\end{equation}
By BCH relations the squeeze operator can be written as 
\begin{eqnarray}
S(z) 
& = &  \exp\left[{\frac{1}{2}}e^{i\phi}(\tanh r)
a^{\dagger}a^{\dagger}\right]
\left({\frac{1}{\cosh r}}\right)^{({{\frac{1}{2}}+a^{\dagger}a})}
\exp\left[-{\frac{1}{2}}e^{-i\phi}(\tanh r)aa\right] \label{b} \\
& = &  \exp\left[{\frac{1}{2}}e^{i\phi}(\tanh r)
a^{\dagger}a^{\dagger}\right]
(\cosh r)^{-1/2}\sum_{n=0}^{\infty}
      \frac{({\rm sech} r -1)^n}{n!}(a^{\dagger})^n(a)^n   \nonumber \\
&~& ~~~~~~~~\times~
\exp\left[-{\frac{1}{2}}e^{-i\phi}(\tanh r)aa\right]~. \label{c}
\end{eqnarray}

The functional forms of $D$ and $S$  are \cite{bchst}
\begin{equation}
D(\alpha) = \exp[-ix_0p_0/2]\exp[ip_0x]\exp[-x_0\partial]~, \label{disp}
\end{equation}
\begin{equation}
S={\cal S}^{-1/2} \exp\left[\frac{iz_2}{2r}\frac{\sinh r}{{\cal S}}(x^2)\right]
\exp[-(\ln {\cal S})(x\partial)]
\exp\left[\frac{iz_2}{2r}\frac{\sinh r}{{\cal S}}(\partial^2)\right]~,
\label{sfunc}
\end{equation}
where 
\begin{equation}
{\cal{S}} = \cosh r + \frac{z_1}{r} \sinh r
=\cosh r + \cos \phi \sinh r
= e^r \cos^2\frac{\phi}{2} +e^{-r}\sin^2\frac{\phi}{2}~, \label{squeeze}
\end{equation} 
and one should recall the operator definitions on a function $h$:
\begin{eqnarray}
\exp[c\partial]h(x) &=& h(x+c)    \label{func1} \\
\exp[\tau(x\partial)] h(x) &=& h(xe^{\tau})  \label{func2} \\
\exp[c(\partial^2)] h(x) &=& 
\frac{1}{[4\pi c]^{1/2}} \int_{-\infty}^{\infty}
\exp\left[-\frac{(y-x)^2}{4c}\right] h(y) dy ~.  \label{func3}
\end{eqnarray}

Using these functional forms for $D$ and $S$
the most general squeezed wave function is \cite{bchst}
\begin{eqnarray}
\psi_{ss} &=& D(\alpha)S(z)\psi_0 \nonumber \\
 &=&
 \frac{1}{\pi^{1/4}}\frac{\exp[-ix_0p_0/2]}{[{\cal S}(1 +i2\kappa)]^{1/2}}
  \exp\left[-(x-x_0)^2 \left(\frac{1}{2{\cal S}^2(1+i2\kappa)}-i\kappa\right)
    +ip_0x\right]~,
\end{eqnarray}
where
\begin{equation}
\kappa \equiv \frac{z_2 \sinh r}{2r{\cal S}}~.
\end{equation}
Setting $z$ to be real and positive,  
yields the most commonly  studied example:
\begin{equation}
  \psi_{ss} =
  [\pi^{1/2}s]^{-1/2}\exp\left[-\frac{(x-x_0)^2}{2s^2} - ip_0x\right]~,
 ~~~ ~~~~s = e^r~.
\end{equation}


\section{The Forgotten Displaced Number States}

In the 1950's, before the modern work on coherent states started, 
there was a short flurry of activity by four authors, which has 
essentially been forgotten.  These papers were by Senitzky \cite{sin} 
(perhaps the best known), Plebanski \cite{pleb}, Husimi \cite{hus}, 
and Epstein \cite{ep}.  These authors investigated 
whether there are other wave packets, besides
the coherent-state packets, 
which keep their shapes and follow the classical
motion.  They found that there are.   In our modern parlance, they found
that any displaced number state  follows the classical
motion and keeps its shape.  

In particular, given that the number-state 
wave functions are 
\begin{equation}
\psi_n =   \exp\left[-\frac{x^2}{2}\right]
  \frac{H_n(x)}{\pi^{1/4}[2^n n!]^{1/2}} ~,
\end{equation}
the displaced number states are 
found by applying the functional form of 
the displacement operator onto the number wave functions: 
\begin{equation}
\psi_{cs(n)}=D(\alpha) \psi_n 
= \frac{\exp[-ix_0p_0/2]\exp[ip_0x]}{\pi^{1/4}}
  \exp\left[-\frac{(x-x_0)^2}{2}\right]
  \frac{H_n(x-x_0)}{[2^n n!]^{1/2}} ~. \label{cns}
\end{equation}

An equivalent result can be obtained in Fock notation by writing  
\begin{eqnarray}
D(\alpha)|n\rangle &=& \exp[-\alpha^2/2]\exp[\alpha a^{\dagger}]
      \exp[-\alpha^* a]|n\rangle \\
&=&\exp[-\alpha|^2/2]\sum_{k=0}^{\infty} \frac{\alpha^k}{k!}
   \sum_{j=0}^{n}\frac{(-\alpha^*)^j}{j!}
   \left[\frac{(n-j+k)!n!}{(n-j)!(n-j)!}\right]^{1/2}|n-j+k\rangle.
\end{eqnarray}
Then by changing from the Fock state basis to the wave function 
basis, interchanging the order of the summations, and 
using the sum  49.4.2 of Ref. \cite{hansen}, one has 
\begin{eqnarray}
D(\alpha)|n\rangle &\rightarrow & \pi^{-1/4}
  \exp[-(x-\sqrt{2}\alpha_1)^2/2 +i\sqrt{2}\alpha_2 x -i\alpha_1\alpha_2]
   \nonumber  \\
&~&~~~\frac{(n!)^{1/2}}{2^{n/2}}\sum_{j=0}^{n} 
       \frac{(-\alpha^*)^j 2^{j/2}}{j!(n-j)!} H_{n-j}(x-\alpha/\sqrt{2}).
\end{eqnarray}
Changing the sum over $j$ to a sum over $k=n-j$, the sum 
is in the form of the seventh equality  on page 255 
of Ref. \cite{mos}.  Using this  one obtains the same answer 
as Eq. (\ref{cns}) above. 

Time-evolution can be shown to be described by \cite{bchst}
\begin{equation}
x_0\rightarrow x_0(t)=(x_0\cos t + p_0\sin t)~, ~~~~~~~
p_0\rightarrow p_0(t)=(p_0\cos t -x_0\sin t)~.
\end{equation}
The wave packets have the properties 
\begin{equation}
[\Delta x(t)]^2 = [\Delta p(t)]^2 = (n+ 1/2)~,
\end{equation}
\begin{equation}
\langle H\rangle = (n+1/2) + |\alpha|^2~.
\end{equation}

In the later excitement over ordinary coherent and squeezed states, 
the work of the earlier authors \cite{sin}-\cite{ep}
was, unfortunately, mainly forgotten.  
In 1973 one paper appeared apllying the operator-formalism 
displacement operator on the Fock number state \cite{cs1}.  
Then, in the 1980's, papers started appearing studying these states
\cite{cs2}-\cite{cs4}, one of which \cite{cs2} noted the early work by 
Senitzky and others \cite{sin}-\cite{ep}.


\section{Squeezed Number States}

Plebanski \cite{pleb}, in wave-function form,  also looked a little into 
what we would call squeezed number states.  By using 
the functional forms in Eqs. (\ref{disp}) and (\ref{sfunc}) 
of $D$ and $S$ \cite{bchst}, we find that the most general squeezed 
number states are 
\begin{eqnarray}
\psi_{ss(n)} &=& D(\alpha)S(z)\psi_n   \nonumber   \\
&=&  \frac{\exp[-ix_0p_0/2]}{\pi^{1/4}[{\cal F}_1]^{1/2}}
\exp\left[-\frac{(x-x_0)^2}{2}{\cal F}_2+ip_0x\right]  \nonumber \\
 &~~~~~~~~~& \left[\left({\cal F}_3\right)^{n/2}
   \frac{1}{[2^nn!]^{1/2}}H_n\left(\frac{x-x_0}{{\cal F}_4}
    \right)\right]~,  \label{sn}
\end{eqnarray}
where
\begin{eqnarray}
{\cal F}_1 &=& {\cal S}(1+i2\kappa)  
         = \cosh r +e^{i\phi}\sinh r,   \label{f1} \\
{\cal F}_2 &=& \left(\frac{1}{{\cal S}^2(1+i2\kappa)}-i2\kappa\right)
   = \left(\frac{1}{{\cal S}^2(1+4\kappa^2)}\right) -i2\kappa 
     \left[ 1 + \left(\frac{1}{{\cal S}^2(1+4\kappa^2)}\right)\right]
      \nonumber  \\       
    &=& \frac{1-i\sin\phi \sinh r(\cosh r + e^{i\phi}\sinh r)} 
           {(\cosh r + \cos\phi \sinh r)(\cosh r + e^{i\phi}\sinh r)},  \\
{\cal F}_3 &=& \frac{1-i2\kappa}{1+i2\kappa}  
         = \left(\frac{\cosh r + e^{-i\phi}\sin{\phi}\sinh r} 
                 {\cosh r + e^{i\phi}\sin{\phi}\sinh r}\right)
                = e^{-i2\tan^{-1}(2\kappa)},  \\
{\cal F}_4 &=& {\cal S}(1+4\kappa^2)^{1/2}  
         = [\cosh^2r + \sinh^2 r + 2\cos\phi \cosh r \sinh r]^{1/2}. 
           \label{f4} 
\end{eqnarray}
By using properties of the ${\cal F}$'s. such as 
that ${\cal F}_3$ is 
a phase, $[{\cal F}_2 + {\cal F}_2^*] = 2/{\cal F}_4^2$, 
and  ${\cal F}_1^*{\cal F}_1 = {\cal F}_4^2$, one can verify that
\begin{equation}
       1= \int_{-\infty}^{\infty}dx~\psi_{ss(n)}^*(x)~\psi_{ss(n)}(x).
\end{equation}
Note that when $n=0$, the final large square bracket of Eq. (\ref{sn})
is unity and the wave function reduces  to the normal squeezed states.
 
Satyanarayana \cite{cs3}  defined the 
problem in Fock notation.  
He obtained that the squeezed number states are
\begin{eqnarray}
|\alpha,z,n\rangle &=& D(\alpha)S(z)|n\rangle  \nonumber \\
 &=&\exp[-|\alpha|^2/2]\sum_{m,l}G_{mn}(z)\left(\frac{m!}{l!}\right)^{1/2}
         L_(m)^{(l-m)}(|\alpha|^2)\alpha^{l-m}|l\rangle~,
\end{eqnarray}
where the $G_{mn}(z)$ are given by infinite sums.
Later discussions followed.  One emphasized number states  having the 
displacement operator and squeeze operator being applied separately
(rather than in combination) \cite{knight}.  Another was interested in
these states from the point of view of the parity operator \cite{vourdas}.
A third still addressed the topic from the viewpoint of molecular wave packets
\cite{wolfie}. 

Now we will show how to sum the Fock problem in closed form, thereby 
reproducing the result of Eq. (\ref{sn}). Start with a number 
state and apply the squeeze operator.  If 
\begin{equation}
d = \frac{1}{2}e^{i\phi}\tanh r,
\end{equation}
then
\begin{eqnarray}
S(z)|n\rangle &=& e^{da^{\dagger}a^{\dagger}}\left(\frac{1}{\cosh r}\right)^
     {\frac{1}{2}+a^{\dagger}a}e^{-d^*aa}|n\rangle   \label{sa}  \\
 &=& e^{da^{\dagger}a^{\dagger}}\left(\frac{1}{\cosh r}\right)^
     {\frac{1}{2}+a^{\dagger}a}\sum_{j=0}^{\left[\frac{n}{2}\right]}
     \left[\frac{n!}{(n-2j)!}\right]^{1/2}
     \frac{(-d^*)^j}{j!}|n-2j\rangle  \label{sb}  \\
&=& \left(\frac{1}{\cosh r}\right)^{n+1/2}(n!)^{1/2}
     \sum_{j=0}^{\left[\frac{n}{2}\right]}
     \frac{(-d^*)^j(\cosh r)^{2j}}{(n-2j)!j!}  \nonumber  \\
&~& ~~~~~~~~  \sum_{k=0}^{\infty}\frac{d^k[(n-2j+2k)!]^{1/2}}{k!}
      |n-2j+2k\rangle, \label{sc}
\end{eqnarray}
where, in the upper limit of the sums, $[n/2]$ is the greatest integer
function.

With the identification
\begin{equation}
|n-2j+2k\rangle \rightarrow \frac{e^{-x^2/2}H_{n-2j+2k}(x)}
          {\pi^{1/4}[2^{n-2j+2k}(n-2j+2k)!]^{1/2}},
\end{equation}
and using the known infinite sum 49.4.4 of Ref. \cite{hansen}, 
\begin{equation}
\sum_{k=0}^{\infty}\frac{v^k}{k!}H_{2k+m}(x) = (1+4v)^{-m/2-1/2}
    \exp\left[\frac{4vx^2}{1+4v}\right]H_m\left(\frac{x}{[1+4v]^{1/2}}\right),
\end{equation}
one finds
\begin{equation}
S(z)|n\rangle \rightarrow \frac{\exp[-x^2{\cal F}_2/2] (n!)^{1/2}}
{\pi^{1/4}{\cal F}_1^{1/2}[\cosh r(\cosh r +e^{i\phi}\sinh r)]^{n/2}2^{n/2}}
R,
\end{equation}
where
\begin{equation}
R = \sum_{j=0}^{\left[\frac{n}{2}\right]}
     \frac{\tau^j}{(n-2l)!j!} H_{n-2j}(xy),    \label{rsum}
\end{equation}
and
\begin{equation}
\tau=-e^{-i\phi}\sinh r(\cosh r +e^{i\phi}\sinh r), ~~~~~~~
    y= \frac{\cosh r}{{\cal F}_1^{1/2}}.
\end{equation}

To do the sum in Eq. (\ref{rsum}), we start 
with the $n$ is even case, $[n/2]= n/2$. By changing
the variable of the sum to $k=n/2-j$ and doing a little algebra, one finds
\begin{equation}
R_E = \frac{\tau^{n/2}}{(n/2)!}\sum_{k=0}^{n/2}
    \frac{(-n/2)_k}{(2k)!}\left(\frac{-1}{\tau}\right)^k H_{2k}(xy).
\end{equation}
But this sum is of the form of equation 49.4.12 of Ref. \cite{hansen}:
\begin{equation}
\sum_{k=0}^{n}
\frac{(-n)_k}{(2k)!}v^k H_{2k}(x)=\frac{n!}{(2n)!}(-1-v)^n
    H_{2n}[x(1+1/v)^{-1/2}].  \label{evensum}
\end{equation}
Using this and a fair amount of further algebra one finally obtains 
for the $n$ being even case ($n=n_E$),
\begin{equation}
S(z)|n=n_E\rangle \rightarrow
\frac{1}{\pi^{1/4}[{\cal F}_1]^{1/2}}
\exp\left[-\frac{x^2}{2}{\cal F}_2\right]  
         \left[\frac{\left({\cal F}_3\right)^{n/2}}
   {[2^nn!]^{1/2}}H_n\left(\frac{x}{{\cal F}_4}
    \right)\right]~.  \label{snb}
\end{equation}
But this is exactly the form of Eq. (\ref{sn}) for the special case of no 
displacement, $x_0=p_0=0$.  Therefore, now applying the displacement operator
trivially yields the general result of Eq. (\ref{sn}), as it should.

The $n$ being odd case $n=n_O$ is obtained in the same manner. The only 
significant difference is that instead of using the sum of Eq. 
(\ref{evensum}), one uses Eq. 49.4.14 of Ref. \cite{hansen}:
\begin{equation}
\sum_{k=0}^{n}
\frac{(-n)_k}{(2k+1)!}v^k H_{2k+1}(x)=\frac{n!}{(2n+1)!}
    (-v)^n(1+1/v)^{n+1/2}
    H_{2n+1}[x(1+1/v)^{-1/2}].  \label{oddsum}
\end{equation}
The identical functional form is obtained and so we have demonstrated,
as it must be, that the functional and operator formalisms yield the 
identical squeezed number states.


\section{Time-Dependent Uncertainties}

Calculating the time-dependent 
uncertainties is most easily done with the 
boson-operator formalism.  This is because we know that  
\begin{eqnarray}
D^{\dagger}(\alpha) a D(\alpha) &=& a - \alpha~, \\
S^{\dagger}(z) a S(z) &=& (\cosh r)a + e^{i\phi}(\sinh r) a^{\dagger}~, \\
T^{\dagger}(t) a T(t) &=& ae^{-it}~,
\end{eqnarray}
where $T$ is the time-evolution operator
\begin{equation}
T(t) = \exp[-iHt] = \exp[-i(a^{\dagger}a + 1/2)t]
      = \exp[-i(-\partial^2 +x^2)t/2]~.  \label{T}
\end{equation}
With this 
it is straight-forward to calculate the uncertainties, as a function of time,
for the squeezed number states.

To begin, 
\begin{equation}
\langle x_s(t) \rangle_n = \langle n|S^{\dagger}D^{\dagger}T^{\dagger}
                         x TDS|n\rangle 
      = x_0 \cos t + p_0 \sin t~.
\end{equation}
Similarly, 
\begin{eqnarray}
\langle p_s(t) \rangle_n &=& p_0 \cos t - x_0 \sin t~, \\
\langle x_s^2(t) \rangle_n &=& (2n+1)
[(\cosh r)^2 + (\sinh r)^2 
+2(\cosh r)(\sinh r)\cos(2t -\phi)] \nonumber \\
  &~& ~~~~~+\langle x_s(t) \rangle_n^2~, \\
\langle p_s^2(t) \rangle_n &=& (2n+1)
[(\cosh r)^2 + (\sinh r)^2 
-2(\cosh r)(\sinh r)\cos(2t -\phi)]  \nonumber \\
   &~&~~~~~+\langle p_s(t) \rangle_n^2~. 
\end{eqnarray}
Therefore, the uncertainty-product as a function of time is 
\begin{eqnarray}
\frac{[\Delta x(t)]^2[\Delta p(t)]^2}{(n+1/2)^2}
&=& 1 + 4(\cosh r)^2(\sinh r)^2\sin^2(2t - \phi) \label{uncert1} \\
&=&   1 +\frac{1}{4}\left(s^2-\frac{1}{s^2}\right)^2\sin^2(2t-\phi)~.
    \label{uncert}
\end{eqnarray}
Eq. (\ref{uncert1}) agrees with the results of Ref. \cite{vourdas} for 
$t=0$ and 
Eq. (\ref{uncert}) agrees with the standard squeezed-state 
result when $n=0$ \cite{qop}.


\section{Time-Evolution of the States}

Taking the functional definition of the time-displacement operator, $T$,
in Eq. (\ref{T}), and using BCH relations  it has been shown \cite{bchst}
that the harmonic-oscillator time-displacement operator 
can be written as 
\begin{equation}
T = [\cos t]^{-1/2}\exp\left[-\frac{i}{2} \tan t(x^2)\right]
\exp[-(\ln \cos t)(x\partial)]
\exp\left[\frac{i}{2}\tan t (\partial^2)\right]~.
\end{equation}
Using this result,
along with the functional definitions in Eqs. 
(\ref{func1}-\ref{func3}), and  combining it with the squeezed state 
wave function, $\psi_{ss(n)}$ of Eq. (\ref{sn}), one has
\begin{equation}
\Psi_{ss(n)}(x,t) = T\psi_{ss(n)}(x)  
      = \frac{e^{-i\tan t~x^2/2}e^{-ix_0p_0/2}{\cal F}_3^{n/2}}
         {\pi^{3/4}[i\sin t {\cal F}_1 2^{n+1} n!]^{1/2}}~ I,
\end{equation}
where 
\begin{equation}
I=\int_{-\infty}^{\infty}dy~\exp\left[-\frac{(y-x/\cos t)^2}{i2\tan t}
      -\frac{1}{2}(y-x_0)^2{\cal F}_2+ip_0y\right]
      H_n\left(\frac{y-x_0}{{\cal F}_4}\right).
\end{equation}
This integral can be evaluated with the aid of  7.374.8 of Ref. \cite{gr}, 
yielding
\begin{eqnarray}
\Psi_{ss(n)}(x,t)  &=& \left[\frac{1}{\pi^{1/4}[B {\cal F}_1]^{1/2}}\right]
         \left[\frac{{\cal F}_3^n A^n}{2^n n!}\right]^{1/2}
         H_n \left(\frac{X(t)}{{\cal F}_4B[A]^{1/2}}\right)  \nonumber  \\
&~&~~~\exp\left[-\frac{x^2}{2}\frac{{\cal F}_2 \cos t +i \sin t}{B}
     +x\frac{(x_0{\cal F}_2+ip_0)\cos t}{B}
     -\frac{x_0^2}{2}\frac{{\cal F}_2 \cos t}{B}\right] \nonumber \\
&~&~~~~~\exp 
       \left[ -\frac{p_0^2}{2}\left(\frac{i \sin t}{B}\right) 
   -x_0p_0\left(\frac{{\cal F}_2\sin t}{B}+\frac{i}{2} \right)
       \right],
\end{eqnarray}
where the ${\cal F}_i$ are given in Eqs. (\ref{f1}) - (\ref{f4}), and 
\begin{eqnarray}
A&=&\left(1-\frac{i2 \sin t}{{\cal F}_4^2 B}\right)
      \left(\frac{B-i2\sin t/{\cal F}_4^2}{B}\right),   \\
B&=&\cos t + i{\cal F}_2 \sin t,  \\
X(t)&=& x -(x_0 \cos t + p_0 \sin t).
\end{eqnarray}

Appropriate limits can be checked.  For the case $t = 0$. one finds  
$\Psi_{ss(n)}(x,t=0) =\psi_{ss(n)}(x)$, the function given in Eq. (\ref{sn}).
Also, for the non-squeezed case of $z=0$ or ${\cal F}_i=1$, one finds the
appropriate time-evolution of displaced number states:
\begin{eqnarray}
\Psi_{ss(n)}(x,t,{\cal F}_i=1)&=&\frac{e^{-i(n+1/2)t}}{\pi^{1/4}[2^n n1]^{1/2}}
        H_n(X(t))\exp\left[-\frac{X^2(t)}{2}\right] \nonumber  \\
&~&\exp[i\{x-(x_0\cos t + p_0\sin t)/2\}(p_0\cos t -x_0 \sin t)].
\end{eqnarray}

The probability density is given by
\begin{eqnarray}
\rho_{ss(n)}(x,t) &=& \Psi_{ss(n)}^*(x,t)\Psi_{ss(n)}(x,t)  \\
      &=&\frac{[AA^*]^{n/2}}{2^n(n!)[\pi BB^*]^{1/2}{\cal F}_4}
        H_n \left(\frac{X(t)}{{\cal F}_4 B^*[A^*]^{1/2}}\right) 
        H_n \left(\frac{X(t)}{{\cal F}_4 B[A]^{1/2}}\right)   \nonumber  \\  
 &~&    \exp\left[-\frac{X^2(t)}{{\cal F}_4^2 BB^*} \right].
\end{eqnarray}
With the aid of integral 2.20.16.2 of Ref. \cite{prud},
\begin{equation}
\int_0^{\infty}dx~e^{-px^2}H_n(bx)H_n(cx) = 
     \frac{2^{n-1} n! \sqrt{\pi}}{p^{(n+1)/2}}
     (b^2+c^3-p)^{n/2}P_n\left(\frac{bc}{\sqrt{p(b^2+c^2-p)}}\right),
\end{equation}
where the $P_n$ are the Legendre functions, it can 
in principle be demonstrated that
the probability density is properly normalized:
\begin{equation}
1 = \int_{-\infty}^{\infty} dx~\rho_{ss(n)}(x,t).
\end{equation}

However, for small $n$ it is easier to verify the 
normalization  on a case by case basis.
The first three probability densities are
\begin{eqnarray}
\rho_{ss(0)}(x,t) &=&\frac{1}{[\pi BB^*]^{1/2}{\cal F}_4} 
     \exp\left[ -\frac{X^2}{BB^*{\cal F}_4^2}\right],  \\
\rho_{ss(1)}(x,t) &=&\frac{2}{[\pi BB^*]^{1/2}{\cal F}_4} 
     \exp\left[ -\frac{X^2}{BB^*{\cal F}_4^2}\right]
       \left(\frac{X^2}{BB^*{\cal F}_4^2}\right),   \\
\rho_{ss(2)}(x,t) &=&\frac{AA^*}{2[\pi BB^*]^{1/2}{\cal F}_4}
      \exp\left[ -\frac{X^2}{BB^*{\cal F}_4^2}\right] \nonumber \\
&~&~~~~~~~~~~~~~~
\left(\frac{2X^2}{B^{*2}A^*{\cal F}_4^2}-1\right)
\left(\frac{2X^2}{B^{2}A{\cal F}_4^2}-1\right).
\end{eqnarray}
$\rho_{ss(0)}(x,t)$ is the probability density 
for the ordinary squeezed state.    $\rho_{ss(1)}(x,t)$ and 
$\rho_{ss(2)}(x,t)$  are $n=1$ and $n=2$ squeezed number states, 
having two and three humps, respectively, in their wave packets.
In Figs. 1-4  we show three-dimensional plots of $\rho_{ss(1)}$ 
vs. $x$ and $t$ for various $n=1$ squeezed  states.  

Fig. 1 looks at the case $x_0 = 8$, $p_0 = 0$, and $z = \ln 2$.
The picket starts with no velocity at maximum displacement.   
It is at its broadest when $t=0,~\pi$, and $2\pi$ for 
$x=8,~-8$, and $8$.  In between the locations where the 
wave packet is broadest, the two humps
are closer together, narrower, and higher in amplitude. In Fig 2, 
we show the case again having  $x_0 = 8$ and  $p_0 = 0$, but 
this time  $z = -\ln 2$.  This basically changes the phase by $\pi/2$.
The narrow, peaked wave packets are when $t=0,~\pi$, and $2\pi$ for 
$x=8,~-8$, and $8$.  The broader and shorter conditions are in between.

Fig. 3 is again an example with $x_0 = 8$ and  $p_0 = 0$, but now 
$z$ is imaginary:  $z = (\ln 2)\exp[i\pi/2]=i\ln 2$.  One can see from the $t=2\pi$ 
case that now we have a phase shift such that neither extreme configuration
is located at the $t=$ an integral number times $\pi$ positions.
Finally, in Fig. 4, we change to a central position and maximum velocity
to the right at $t=0$:  $x_0 = 0$ and  $p_0 = 8$, with $z = \ln 2$.  
Now the broadest wave 
packets are centered at $x=0$ for $t=0,~\pi$, and $2\pi$.  The narrow 
peaked packets are when $t=\pi/2$ and $3\pi/2$.


\section{Discussion}

There is hope that, in the not too distant future, displaced and
squeezed number states can be observed.  This optimism is based
on the recent work of Wineland's group with trapped $Be^+$ ions
\cite{wine1,wine2}. 

With laser cooling and a series of laser pulses to entangle the electronic 
and motional states they produced (displaced) even and odd coherent 
states \cite{wine1,vogel}.  They also have been able to produce a squeezed 
(but not displaced) ground state and number states \cite{wine2}.  
If all these
techniques can be combined then in principle displaced and squeezed number
states perhaps can be produced.  However, this would be no easy feat since 
there are very complicated heating problems involved.

When these states are produced, they should mimic the features displayed
in our Figs. 1 to 4.


\section*{Acknowledgments}

I gratefully acknowledge conversations 
with Iwo and Sophie Bialynicki-Birula, 
Peter Knight, Wolfgang Schleich, and 
especially with A. Vourdas which took place  at the Humboldt 
Foundation Workshop on Current Problems in Quantum Optics, organized by 
Prof. Harry Paul.  This work was supported by the U.S. Department of 
Energy and the Alexander von Humboldt Foundation.

\baselineskip=.33in

\newpage
\large
\noindent{{\bf Figure Captions}}
\normalsize
\baselineskip=.33in

Figure 1.  A three-dimensional plot of the $n=1$ squeezed-state probability
density $\rho_{ss(1)}$ as a function of position, $x$, and time, $t$, for 
$x_0 = 8$, $p_0 = 0$, and $z = \ln 2$.  The spikes near the maxima of 
$\rho_{ss(1)}$ are an artifact of the numerical routine.  The true, smooth 
maxima follow the tops of the spikes.

Figure 2.  A three-dimensional plot of the $n=1$ squeezed-state probability
density $\rho_{ss(1)}$ as a function of position, $x$, and time, $t$, for 
$x_0 = 8$, $p_0 = 0$, and $z = -\ln 2$.

Figure 3.  A three-dimensional plot of the $n=1$ squeezed-state probability
density $\rho_{ss(1)}$ as a function of position, $x$, and time, $t$, for 
$x_0 = 8$, $p_0 = 0$, and $z = (\ln 2)\exp[i\pi/2]=i\ln 2$.  
The spikes near the maxima of 
$\rho_{ss(1)}$ are an artifact of the numerical routine.  The true, smooth 
maxima follow the tops of the spikes.

Figure 4.  A three-dimensional plot of the $n=1$ squeezed-state probability
density $\rho_{ss(1)}$ as a function of position, $x$, and time, $t$, for 
$x_0 = 0$, $p_0 = 8$, and $z = \ln 2$.


\end{document}